\documentclass[prl,preprint,nofootinbib,floatfix]{revtex4}

\usepackage{amsmath,amsfonts,amssymb,amstext}
\usepackage{graphicx}
\usepackage[colorlinks,linkcolor=blue]{hyperref}
\usepackage{multirow}
\usepackage[matrix,frame,arrow]{xy}

\newcommand{\ket}[1]{\left|#1\right\rangle}
\newcommand{\bra}[1]{\left\langle#1\right|}

\DeclareMathOperator{\Tr}{Tr}

\def\({\left(}
\def\){\right)}

\newcommand{\be}{\begin{equation}}
\newcommand{\ee}{\end{equation}}
\newcommand{\bea}{\begin{eqnarray}}
\newcommand{\eea}{\end{eqnarray}}
\newcommand{\bF}{\begin{figure}}
\newcommand{\eF}{\end{figure}}

\newcommand{\bi}{\begin{itemize}}
\newcommand{\ei}{\end{itemize}}

\begin{document}
\title{Generation and detection of NOON states in superconducting circuits}
\date{\today}

\author{Seth T. Merkel}
\affiliation{Institute for Quantum Computing and Department of Physics and Astronomy, University of Waterloo, Waterloo, ON N2L 3G1, Canada}
\author{Frank K. Wilhelm}
\affiliation{Institute for Quantum Computing and Department of Physics and Astronomy, University of Waterloo, Waterloo, ON N2L 3G1, Canada}

\begin{abstract}
NOON states, states between two modes of light of the form $|N,0\rangle+e^{i\phi}|0,N\rangle$ allow for super-resolution interformetry. We show how NOON states can be efficiently produced in circuit quntum electrodynamics using superconducting phase qubits and resonators. We propose a protocol where only one interaction between the two modes is required, creating all the necessary entanglement at the start of the procedure.  This protocol makes active use of the first three states of the phase qubits.  Additionally, we show how to efficiently verify the success of such an experiment, even for large NOON states, using randomly sampled measurements and semidefinite programming techniques. 
\end{abstract}

\maketitle

\section{Introduction}
The particle-wave duality of quantum physics has striking consequences on the physics of electromagnetic radiation, necessitating a photonics description of electromagnetic radiation, spanning frequencies from $\gamma$ quanta via visible light to microwaves fields.  The phenomena of classical optics are fully characterized by only a tiny fraction of the wealth of states attainable in the quantum world.  Creating and quantifying these others, the {\em nonclassical states} of light, has defined the field of quantum optics \cite{Glauber63,Cahill69}.   

Recently, this quantum-optical paradigm has been extended to the field of circuit quantum electrodynamics (cQED) \cite{Blais04,Schoelkopf08}  where nonlinear superconducting circuits \cite{Insight} are used to manipulate the quantum state of microwaves in waveguides in much the same way as atoms are used to manipulate optical fields in traditional quantum optics.  Unlike atoms, however, superconducting circuits can be tuned over a broader range of parameters, allowing for highly flexible control.  The current state of the art allows for the full control of quantum states up to 12 photons \cite{Hofheinz09} as well as violations of locality in a Bell experiment \cite{Ansmann09,Chow09}.  Small quantum information algorithms have also been successfully demonstrated \cite{DiCarlo09b}.  

A new frontier in this program is the realization of nonclassical states that are entangled between modes in distinct waveguides. A prominent example is the NOON state \cite{Boto00,Kok02,Sanders89}
\begin{equation}
\ket{\psi}=\frac{1}{\sqrt{2}}\left(\ket{N,0}+e^{i\phi}\ket{0,N}\right),
\end{equation} 
i.e., an equal coherent superposition of $N$ photons in the first mode and vacuum in the second with the reverse, vacuum in the first and $N$ photons in the second.  NOON states were introduced in quantum metrology as a means of surpassing the shot noise limit for phase sensitivity achievable by classical light by a factor of $1/\sqrt{N}$, which instead saturates the Heisenberg bound.  At optical frequencies, NOON states up to $N=4$  have been realized to date,  relying on cascaded pair sources and postselection \cite{Mitchell04,Nagata07}. Schr\"odinger cat states interpreted as NOON states up to $N=10$ have been realized in spin systems, along with an enhanced sensitivity to the magnetic field \cite{Jones09}. 

We propose a scheme to create NOON sates in cQED. In cQED, sources that create pairs of photons {\em independent} of the number of photons already in the mode are currently not available. These would be needed for direct implementation of schemes developed at optical frequencies. We will circumvent this restriction by utilizing an additional energy level in the superconducting qubit we are considering for our artificial atom.  We also place bounds on the size of NOON state that can be achieved and propose a scheme how to verify the preparation of a NOON state using single-qubit measurements. 

\section{System and Hamiltonian}

\begin{figure}
\includegraphics[width=0.9\columnwidth]{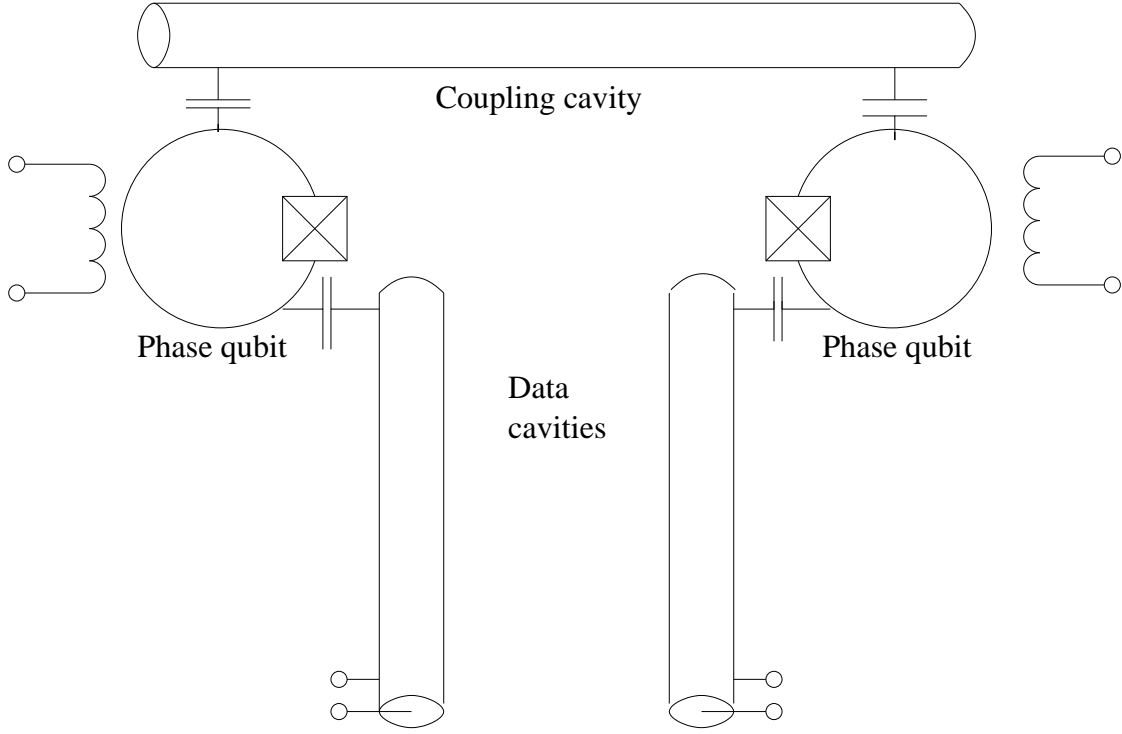}
\caption{Schematic of an experimental setup for generating NOON states consisting of two phase qubits, a coupling cavity, and two data cavities. The NOON state will be between the data cavities, which in the end are  disentangled from the rest of the system. The boxed $\times$'s label Josephson junctions and the small circles external ports to which driving fields can be applied\label{fig:2cavity}}
\end{figure}

The system under consideration consists of two microwave cavities, each coupled to its respective phase qubit. The qubits are also coupled to each other, either directly \cite{Berkley03,McDermott05,Steffen06} or by a coupling cavity \cite{Sillanpaa07,Majer07}. This being a non-essential difference, we are going to focus on the case of cavity-mediated interaction, where by programming the frequency in phase qubits one achieves tuneability .  A schematic this system is given in Fig. \ref{fig:2cavity}, and is described by the Hamiltonian
\be
\hat{H}(t) = \hat{H}_0 + \hat{H}^c_{Q1}(t)+\hat{H}^c_{Q2}(t)+\hat{H}^c_{C1}(t)+\hat{H}^c_{C2}(t).
\ee

The static (or 'drift')  part of the Hamiltonian can be decomposed as
\be
\hat{H}_0 = \sum_{i=1,2}\hat{H}_{Qi} + \sum_{j=1,2,C}\hat{H}_{Cj} + \sum_{i,j}\hat{H}_{QiCj}.
\ee 
Here, $\hat{H}_{Qi}=E_{1i} \ket{1}_i\bra{1}_i+E_{2i} \ket{2}_i\bra{2}_i$ is the Hamiltonian for the phase qubit truncated to its three lowest energy levels $\ket{0}$, $\ket{1}$, and 
$\ket{2}$.  Higher states will never be resonant during the whole protocol. The relevant cavity modes are described by $\hat{H}_{Cj}=\omega_j\hat{a}_j^{\dagger} \hat{a}_j$ with $j=1,2,C$ and $\hat{a}_j^\dagger$ and $\hat{a}_j$ being the photon raising and lowering operators respectively.  With the same index space, the qubit-cavity couplings can be written in rotating wave approximation (RWA), assuming $g_{\rm ij}^{(k)}\ll \omega_j,E_{1i}$ as
\begin{equation}
\hat{H}_{QiCj}= \left( g_{ij}^{(1)} \ket{1}_i\bra{0}_i+g_{ij}^{(2)} \ket{2}_i \bra{1}_i \right) \hat{a}_j +{\rm h.c.}
\end{equation}
which is a straightforward three-level generalization of the Jaynes-Cummings (JC) interaction \cite{Jaynes63}.

The system is primarily manipulated by controlling the phase qubits.  Driving one of the qubits with an external field results in a control of the form, $\hat{H}^{c}_{Qi}(t)=\frac{\Omega_i(t)}{2}\left(\ket{0}_i\bra{1}_i+\lambda \ket{1}_i\bra{2}_i+\textrm{h.c.}\right)$.  Here, $\lambda\approx \sqrt{2}$ is a constant whose precise value depends on details of the qubit. Generally, $\Omega(t)$ will be a resonant microwave drive $\Omega(t) = \Omega_0 \cos(\omega_d t)$.  Additionally, changing the bias current gives a time-dependent $E_{1i}$ and $E_{2i}$ approximately following $E_{1i}=pE_{2i}$ with a fixed $p<2$.  Changing the energy of the qubit will primarily be used as a switch to move interactions in and out of resonance.  Finally, the cavities themselves can be driven externally by microwave sources $H^{(c)}_{Ci}=\lambda_i(t)(\hat{a}_i^\dagger+\hat{a}_i)$.  

\section{NOON state preparation}
\subsection{Coupling and decoupling}
\begin{figure}
\includegraphics[width=0.9\columnwidth]{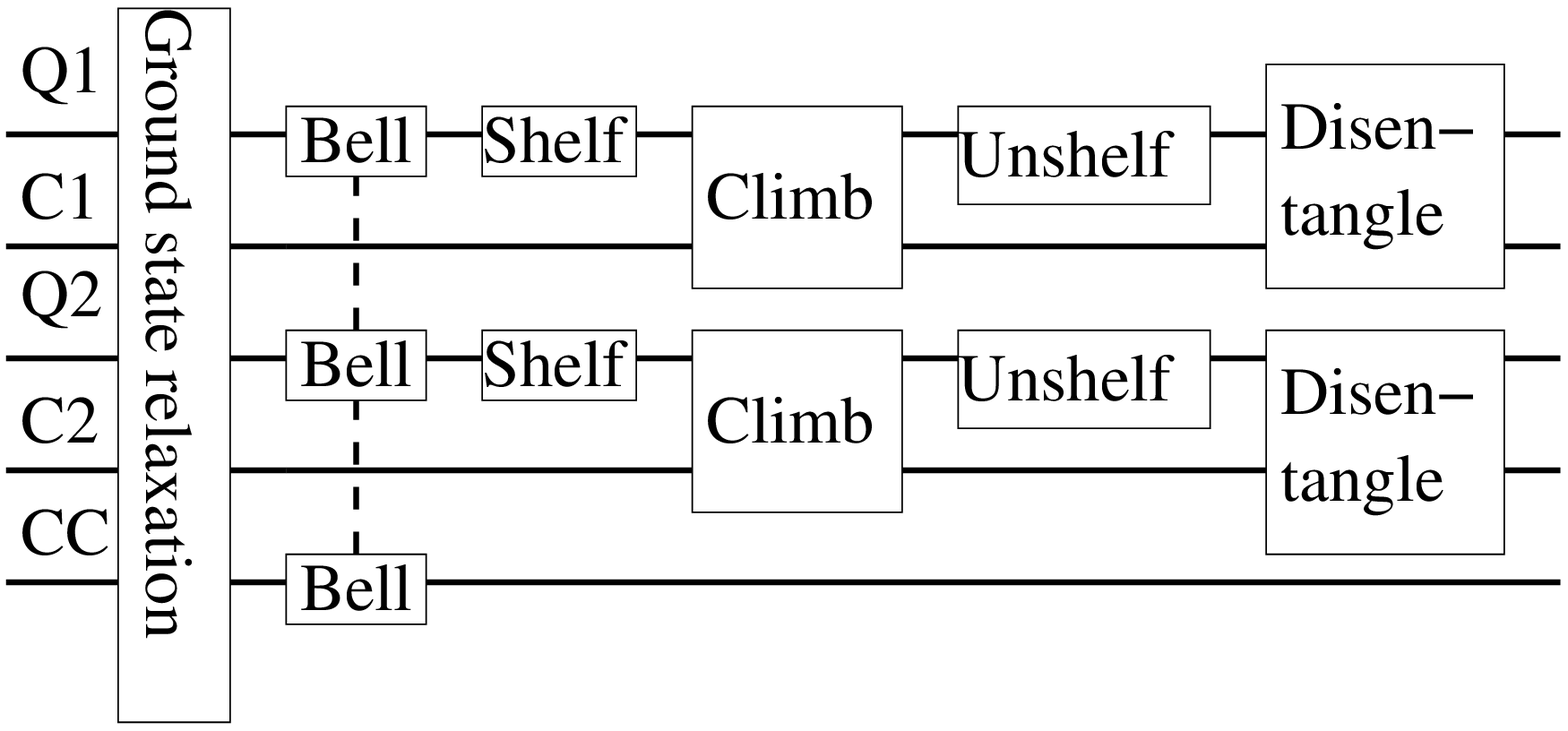}\\
\includegraphics[width=0.4\columnwidth]{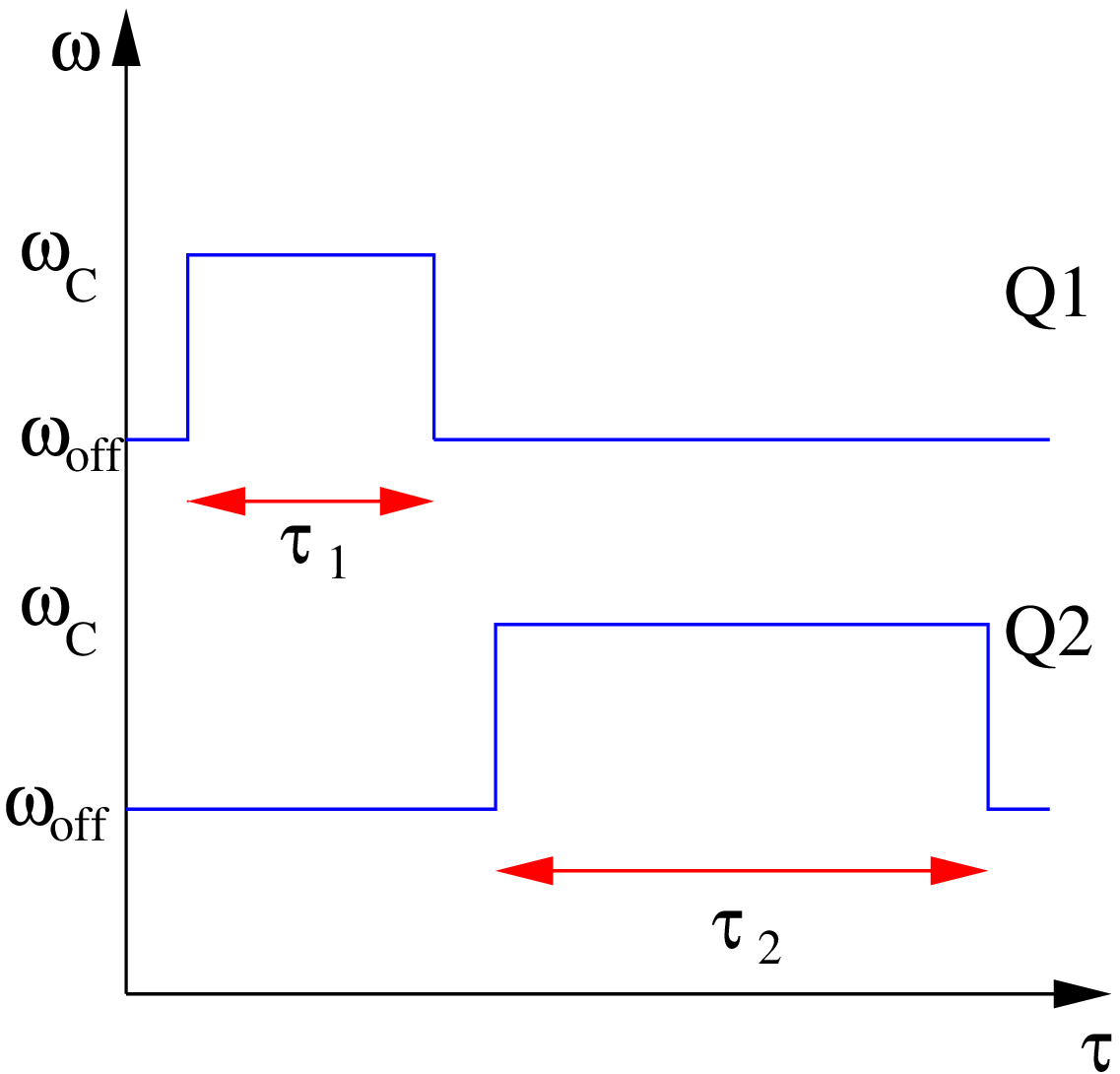}\quad
\includegraphics[width=0.4\columnwidth]{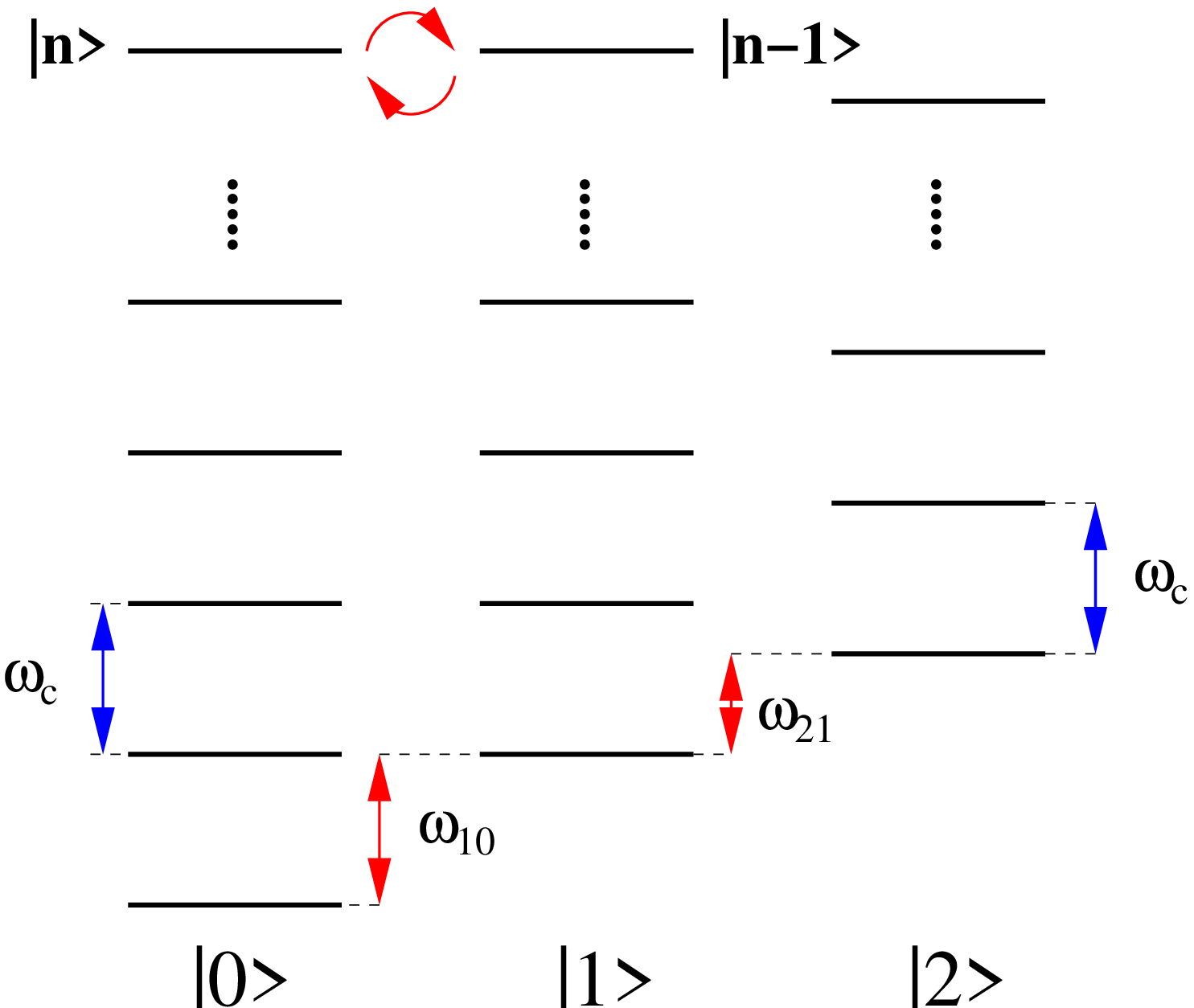}
\caption{The NOON-state preparation protocol. Top: Overview over the steps, highlighting qubits 1 and 2 and cavitites 1, 2 and C. Parties 1 and 2 only interact during the initial Bell state preparation. Bottom left: Bell-state preparation by interacting qubits 1 and 2 with the cavity.  Qubit 1 has already undergone a $\pi$-rotation and is in the state $\ket{0}+\ket{1} / \sqrt{2}$.  The time axis shows interation times $\tau_1$ and $\tau_2$ necessary for performing a $\sqrt{\rm iSWAP}$ operation with the first and an iSWAP with the second qubit. The frequency axis indicates that the interaction is controlled by moving the qubits in and out of resonance with the coupling cavity. Bottom right: Energy scheme indicating the cavity energy ladders for the three different phase qubit states, assuming resonance between the cavity and the $0\leftrightarrow 1$ transition of the qubit. In the last 'unshelving' step of the protocol, $|0,n\rangle$ and $|1,n-1\rangle$ are swapped.\label{fig:protocol}} 
\end{figure}

In order to understand the coupling Hamiltonian, it is helpful to transform to the interaction representation with respect to the $\hat{H}_{Qi}$ and $\hat{H}_{Cj}$. In that frame
\begin{equation}
\hat{H}^I_{QiCj}=\left(g_{ij}^{(1)} \ket{1}_i\bra{0}_i e^{i\Delta_{ij}^{(1)}t}+g_{ij}^{(2)} \ket{2}_i\bra{1}_i e^{i\Delta_{ij}^{(2)}t}\right)
\hat{a}_j +{\rm h.c.}.
\end{equation}
with the detuning $\Delta_{ij}^{(k)}=E_i^{(k)}(t)-E_i^{(k-1)}(t)-\omega_j$.  If  $\Delta_{ij}^{(k)}(t) \gg g^{(k)}_{ij}\sqrt{n_j+1}$, a standard rotating wave argument allows us to neglect the terms that couple $\ket{k}_i\ket{n}_j\leftrightarrow \ket{k-1}_i\ket{n+1}_j$.  The implication is that by rapidly tuning $E_i^{(k)}$, one can switch the coupling between the cavity and the qubit on or off.  This has been successfully demonstrated in a multitude of experiments \cite{Sillanpaa07,Hofheinz08,Hofheinz09}.  The two regimes of interest for the following protocol are the case when both detunings are large, leading to uncoupled systems, and the case when $\Delta_{ij}^{(1)} = 0$ and $\Delta_{ij}^{(2)} \gg g^{(2)}_{ij}\sqrt{2}$, which reduces to the standard resonant JC interaction between states 
$\ket{0}_i , \ket{1}_i$ and the qubit only, see fig. \ref{fig:protocol}, lower right.
\begin{equation}
\hat{H}^I_{QiCj}=g_{ij}^{(1)} \left(\ket{1}_i\bra{0}_i \hat{a}_j +\ket{0}_i\bra{1}_i\right).
\end{equation}

\subsection{Bell pairs}

Preparation of Bell states in this setting has already been demonstrated in Refs. \cite{Ansmann09,Chow09}. It works as outlined in fig. \ref{fig:protocol}, lower left using the notation $\ket{q_1,n_C,q_2}$ for qubit 1, cavity C, and qubit 2. Starting from all elements decoupled by detuning, we let the system relax into its ground state $\ket{0,0,0}$. A resonant $\pi$-pulse on $\Omega_1$ generates the state $\ket{1,0,0}$.  Then, we bring qubit $1$ into resonance with the coupling cavity, $\Delta_{1C}^{(1)}=0$, for a time $\tau_1$ such that $g_{1C}^{(1)}\tau_1/2=\pi/2$ to create
$(\ket{1,0,0}-i\ket{0,1,0})/\sqrt{2}$. Next, we move qubit $1$ out of resonance and qubit $2$ into resonance for $g_{2C}^{(1)}\tau_2/2=\pi$ leading to the state $(\ket{1,0,0}+i\ket{0,0,1})/\sqrt{2}$, which is a Bell state between qubits 1 and 2, decoupled and disentangled from the cavity, see fig. \ref{fig:protocol}. After this step, there is no requirement to couple the qubits again throughout the whole protocol, and so we have no further use for the coupling cavity.  All that remains is to separately generate a map on each subsystem, (qubit+cavity)$_i$ denoted by $\ket{q_i,n_i}$, that transforms $\ket{0,0}_i\rightarrow \ket{0,N}_i$ and $\ket{1,0}_i\rightarrow \ket{0,0}_i$. All necessary entanglement has already been created, since an arbitrary NOON state has exactly the same amount of entanglement as a Bell pair. This is entanglement between the sub{\em systems} (i.e. the two cavity/qubit subsystems), as opposed to an entanglement measure that treats the photons themselves as particles that could be made distinguishable by later splitting them into different modes \cite{Boto00,Kok02}. 

\subsection{Shelving}

In order to selectively increase the photon number of the qubit ground-state branch of the superposition while keeping the excited-qubit branch in vacuum, we shelve the excited state $\ket{1,0}_i$ in the second excited state $\ket{2,0}_i$ by selectively driving the qubits at frequency $\hbar\omega_d=E_{2,i}-E_{1,i}$.  The $\ket{2,0}_i$ state is considered `shelved' in the sense that the Jaynes-Cummings (JC) interaction that is resonant with the transition $\ket{0,1}_i\leftrightarrow\ket{1,0}_i$ is far detuned form the transition $\ket{1,1}_i\leftrightarrow\ket{2,0}_i$ as long as $E_{2,i}-2E_{1,i}\gg g_{ii}^{(2)}$, see fig. \ref{fig:protocol}.  The $\ket{2,0}_i$ state is also left invariant by rotations on the first two qubit levels assuming that the nonlinearity leaves it far off-resonant, or through the use of pulse shaping techniques such as DRAG \cite{Motzoi09}. 

\subsection{Climbing the Jaynes-Cummings ladder}

As a next step, the non-shelved state $\ket{0,0}_i$ is transformed into the state $\ket{0,N-1}_i$. This is a special case of the Eberly-Law protocol \cite{Law96} that has been demonstrated in refs \cite{Fink08,Hofheinz08}.  Each step from $\ket{0,n-1}_i$ to $\ket{0,n}_i$ is realized by first exciting the qubit by a resonant $\pi$-pulse, generating $\ket{1,n-1}_i$.  Next, the qubit  is moved into resonance with its cavity for time $g_{ii}^{(1)}\sqrt{n} t/2=\pi$, moving the excitation from the qubit to the resonator, 
see fig. \ref{fig:protocol}.  This is repeated until $\ket{0,N-1}_i$ is reached. At the same time, due to efficient shelving, the other branch of the superposition stays in $\ket{2,0}_i$ as the $\ket{2,0}_i\leftrightarrow \ket{1,1}_i$ transition is off-resonant with the cavity, see fig. \ref{fig:protocol}, lower right.     

\subsection{Unshelving}

The procedure so far brings the total state, written in notation $\ket{n_1,n_2,q_1,q_2,n_C}$ in state $(\ket{N-1,0,0,2,0}-\ket{0,N-1,2,0,0})/\sqrt{2}$.  At this point it may be counter-intuitive as to why we stopped at $N-1$ photons in the previous section, but the simple explantation is that to unentangle the qubits from the cavities we must use the JC-interaction which will inject an additional photon into our state.  The unshelving procedure is as follows.  Applying a $\pi$-pulse to both qubits on the  $\ket{2}_i\leftrightarrow \ket{1}_i$ transition, followed by $\pi$-pulses on the transitions $\ket{1}_i\leftrightarrow \ket{0}_i$ permutes the qubit states such that $\ket{2}_i\rightarrow \ket{0}_i$ and $\ket{0}_i \rightarrow \ket{1}_i$.  This results in the state $(\ket{N-1,0,1,0,0}-\ket{0,N-1,0,1,0})/\sqrt{2}$.  

\subsection{Disentangling}

Now we couple both qubits' to their respective cavities, $\Delta_{ii}^{(1)}=0$, for time $g_{ii}^{(1)}\sqrt{N} t/2=\pi$.  If both the cavity and the qubit are in their ground states they are left invariant by the JC interaction.  On the other hand, the branch of the wavefunction that contains $\ket{1,N-1}_i$ becomes $\ket{0,N}_i$. This step disentangles the cavities from the qubit and leads to the desired NOON-state
\begin{equation}
\ket{\psi}=\frac{1}{\sqrt{2}}\left(\ket{N,0}-\ket{0,N}\right)\otimes\ket{0,0,0}.
\end{equation}

\section{Impact of decoherence, maximal size attainable}

While there is no principal bound on the maximum size $N$ of NOON state this protocol can achieve, there are practical limitations. One will be related to the precision of the control pulses that can be applied, which in principle can be very high using pulse shaping \cite{Khaneja05,Spoerl07,Motzoi09}. More fundamentally, decoherence limits the size of the state attainable. There are two crucial limitations: One is that the qubit state has to be shelved in the second excited state for a long time and the other is that the superposition should not dephase. Given that his happens if any of the two qubits decays, we can go up to $T_2/2$.  The total duration of our protocol can be simply calculated as, 
\begin{equation}
T_{\rm tot}=\frac{1}{g}\sum_{n=1}^{N}\sqrt{\frac{1}{n}}+NT_{\rm Rabi},
\end{equation}
where $T_{\rm Rabi}$ is the time for a single qubit Rabi oscillation. Assuming 
$T_2 \simeq 200 ns$ \cite{Martinis09c} and $T_{\rm Rabi}\simeq 10 ns$ and $1/g\simeq 20 ns$, this allows NOON states up to $N=4$.  These numbers fall well below the number of photons already stored and investigated in such resonators, so photon loss will not be a limiting factor \cite{Hofheinz09,Wang09}. In transmons, which also offer an accessible third level \cite{Koch07c}, coherence times  are longer \cite{Houck09} and larger NOON states may be attainable.

\section{State verification}

Decoherence and imperfect control ensure that in the laboratory setting, the prepared state described by a density matrix $\hat{\rho}_{\rm exp}$  is not a perfect NOON state.  The fidelity between the desired and actual state can be obtained by measuring the overlap 
\begin{equation}
\mathcal{F}=\bra{\psi_{\rm NOON}} \rho_{\rm exp} \ket{\psi_{\rm NOON}}.
\label{eq:fidelity}
\end{equation}
  This is not a trivial task with the measurement tools at our disposal, which consist of driving the system through some simple dynamics followed by qubit measurements.  In this section, we will show that we can create a set of measurement operators that is informationally complete, that is they span the operator space $\mathfrak{su}(d)$ where $d$ is the dimension of the problem under consideration. Here $d=(N+1)^2$ when we consider just the microwave cavities.  The outcomes from an informationally complete set of measurements can be used to perform tomography on the quantum state, but we will also present a method of bounding the fidelity directly.   

Restricting our attention to a single cavity, the types of measurements we consider consist of driving the cavity by a coherent displacement $\alpha$, switching on the JC Hamiltonian for a time $g_{ii}^{(1)} t/2=\tau$, and then measuring the qubit in the $z$-basis by measuring the switching probability out of state $\ket{1}$ \cite{Martinis09c,Insight}.  When we make the simplifying assumption that the qubit is initially uncoupled from the cavity and in its ground state, we can express this observable on the microwave cavity as 
\be
\hat{\mathcal{M}}(\alpha, \tau) = \sum_n \cos(2 \sqrt{n}\tau) \hat{D}^\dagger(\alpha) \ket{n}\bra{n}\hat{D}(\alpha).
\ee  
By sampling different values of the dimensionless time $\tau$, with the amplitude set to $\alpha=0$, one can only infer the diagonal elements of the density operator $\bra{n} \rho  \ket{n}$.  The injection of $\alpha$ is crucial because diagonal measurements alone do not distinguish a NOON state from the mixed state $\hat{\rho}=\frac12\left(\ket{N,0}\bra{N,0} + \ket{0,N}\bra{0,N}\right)$, the latter being useless for quantum metrology.   
In \cite{Hofheinz09}, these types of measurements were used to construct the Wigner function of the cavity, which is a complete description of the quantum state of the cavity mode.  This implies, that at least to some finite truncation of the cavity, the set of measurements $\hat{\mathcal{M}}(\alpha, \tau)$ are informationally complete in that one can can construct a set of $(N+1)^2-1$ linearly independent observables spanning the space $\mathfrak{su}(N+1)$.  To generate such a set it is sufficient to simply sample pairs $(\alpha_j,\tau_j)$ from some random distribution.  This will almost surely generate a linearly independent set of measurement operators.  It is not impossible that there are optimized choices of operators that are close to orthonormal, thus optimizing the sampling of phase space, but in practice it's most often easier to instead just oversample from our distribution on $(\alpha,\tau)$.    

 Correlating the results of measurements from the two individual cavities yields the set of operators $ \hat{\mathcal{M}}_2(\alpha, \tau,\alpha', \tau') = \hat{\mathcal{M}}(\alpha, \tau) \otimes \hat{\mathcal{M}}(\alpha', \tau')$.  This is a complete set of operators on the joint space, however, since the Hilbert space of the two cavity system has dimension $(N+1)^2$, reconstructing the entire density matrix requires measuring at least $(N+1)^4 -1$ of these operators.  To perform full tomography we repeat each measurement a number of times to obtain measurement statistics 
\be
M_j = \Tr \left( \hat{\mathcal{M}}_2(\alpha_j, \tau_j,\alpha'_j, \tau'_j) \rho_{exp} \right) + \sigma_j W_j,
\ee   
where the final term is a Gaussian white noise variable with variance $\sigma_j$.  In principle, one can arbitrarily decrease the strength of this noise by repeated measurements, however, it is generally more efficient to fix $\sigma_j$ and instead sample more measurement operators from our distribution on $\hat{\mathcal{M}}(\alpha_j, \tau_j,\alpha'_j, \tau'_j)$.  From such a measurement record one can obtain a maximum likelihood estimate of the quantum state through a least squares fit \cite{Paris2004}.  The resulting state may have unphysical, negative eigenvalues which need to be compensated for by either simply setting them to zero or by using slightly more involved semi-definite programming techniques \cite{vandenberghe96, Merkel2010} to find the physical state that is most consistent with the measurement outcomes.  The assumption that the qubits were initially in their ground states was not essential and can be circumvented by adding to our set of measurements by additionally performing some unitary operations to the superconducting qubit immediately prior to measurement.  This increases the number of independent measurements necessary for tomography to $16(N+1)^4 -1$ for qubits, or $81(N+1)^4 -1$ when we also account for the second qubit excited state.   

Placing bounds on the experimental value of $\mathcal{F}$ can be accomplished with a smaller set of measurement outcomes.  We can lower bound $\mathcal{F}$, defined in eq. (\ref{eq:fidelity}), by directly solving the optimization problem
\be
\min_\rho \mathcal{F}
\ee
subject to
\be
M_j =\Tr \left( \hat{\mathcal{M}}(\alpha_j, \tau_j,\alpha'_j, \tau'_j) \rho\right),\quad
\rho \geq 0.
\ee
where we have neglected the noise on the measurements.  The second constraint ensures that $\rho$ is a physical density matrix, and thus we will find the state with the lowest overlap with the NOON state that nevertheless agrees with the outcomes from our measurements.  This optimization is of the form of a semidefinite program and can be solved very efficiently numerically \cite{vandenberghe96}.  We used the matlab program yalmip running the solver SeDuMi.   Additionally, since the objective of this optimization ($\mathcal{F}$, eq. \ref{eq:fidelity}) is linear, solving the maximization problem (i.e. finding an upper bound), is of the same difficulty.

\begin{figure*}[t!]
\begin{center}
\begin{tabular}{cc}
(a)&(b)\\
\includegraphics[width=8.7cm,clip]{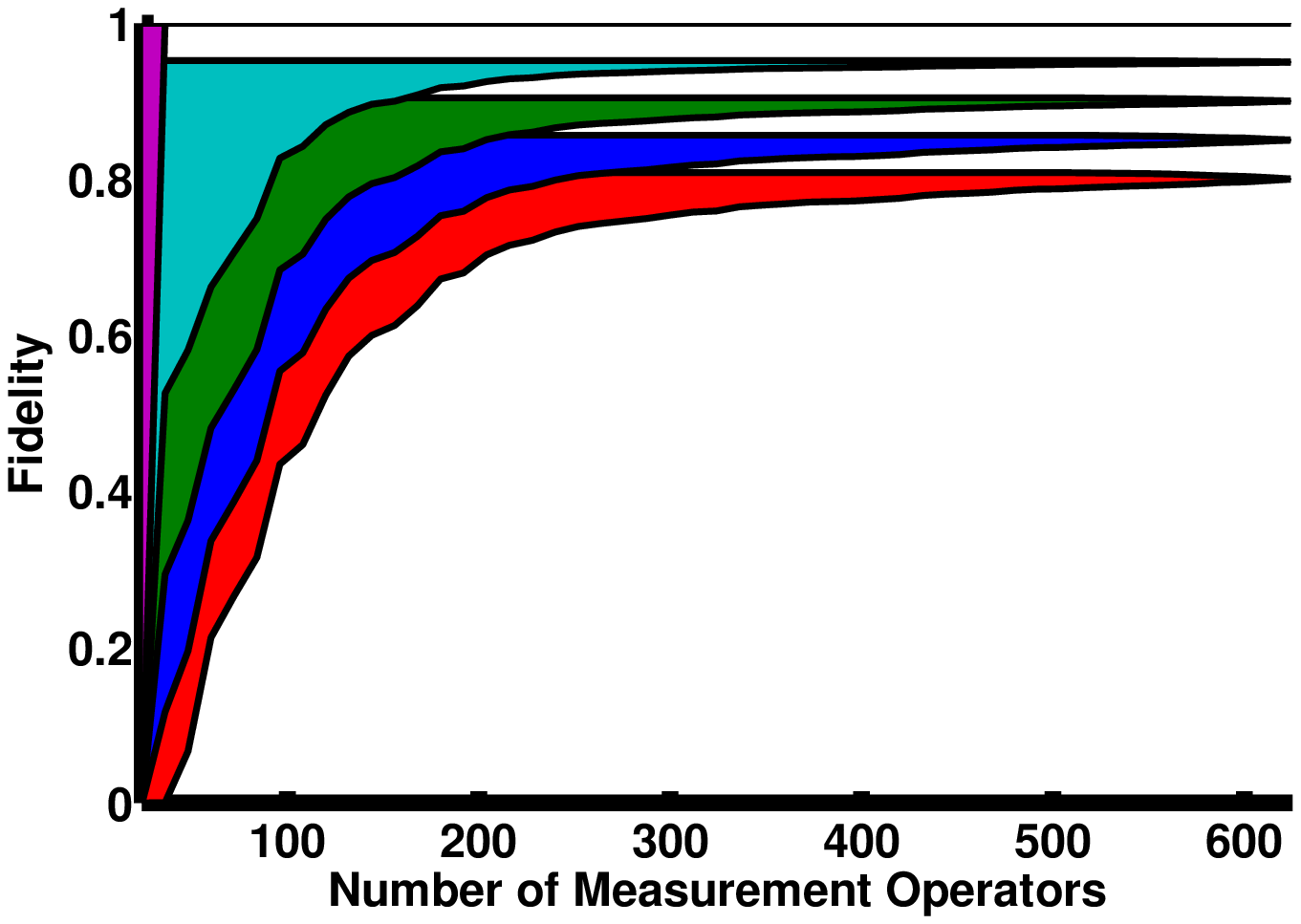} &
\includegraphics[width=8.3cm,clip]{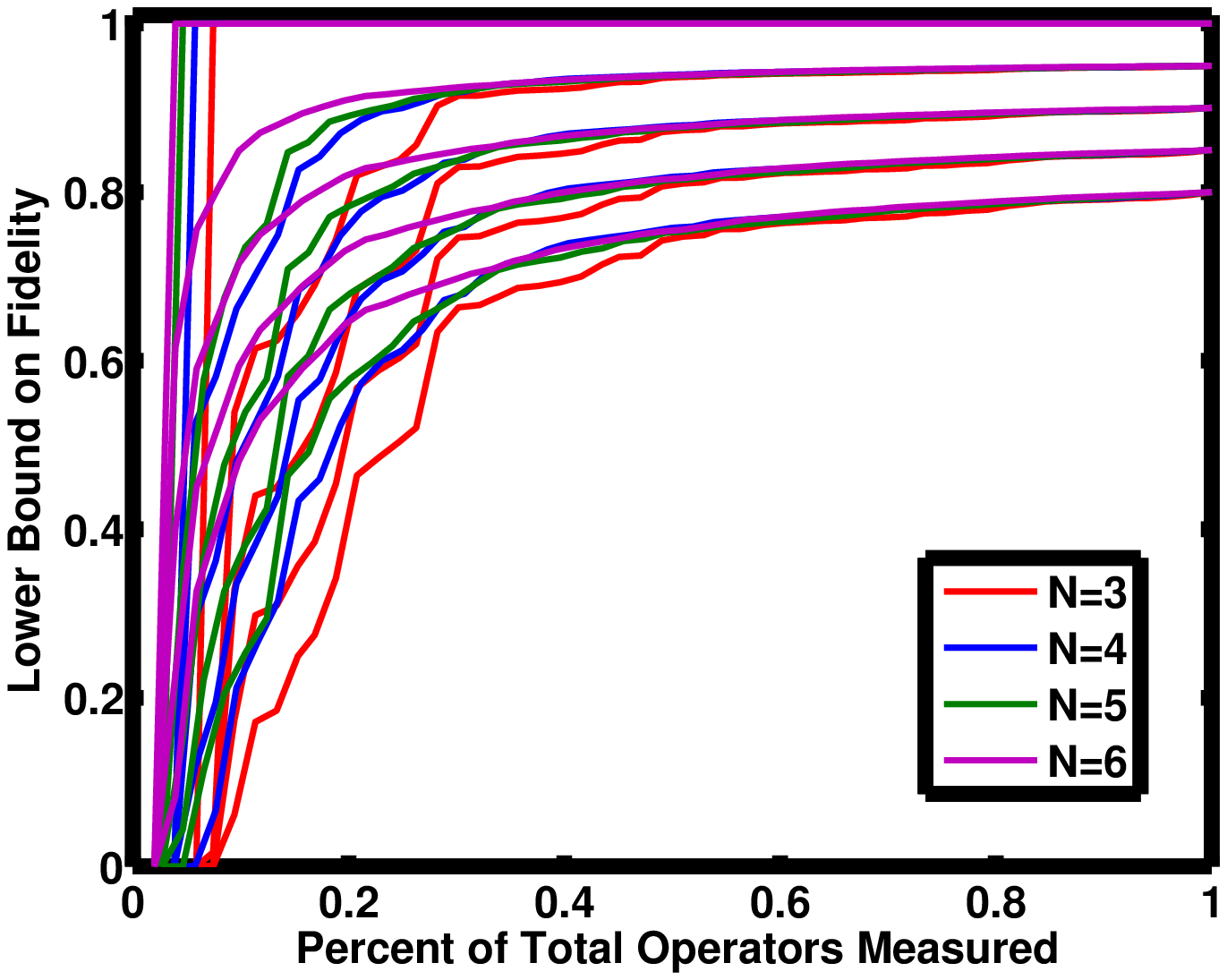} \\
(c)&(d)\\
\includegraphics[width=8.3cm,clip]{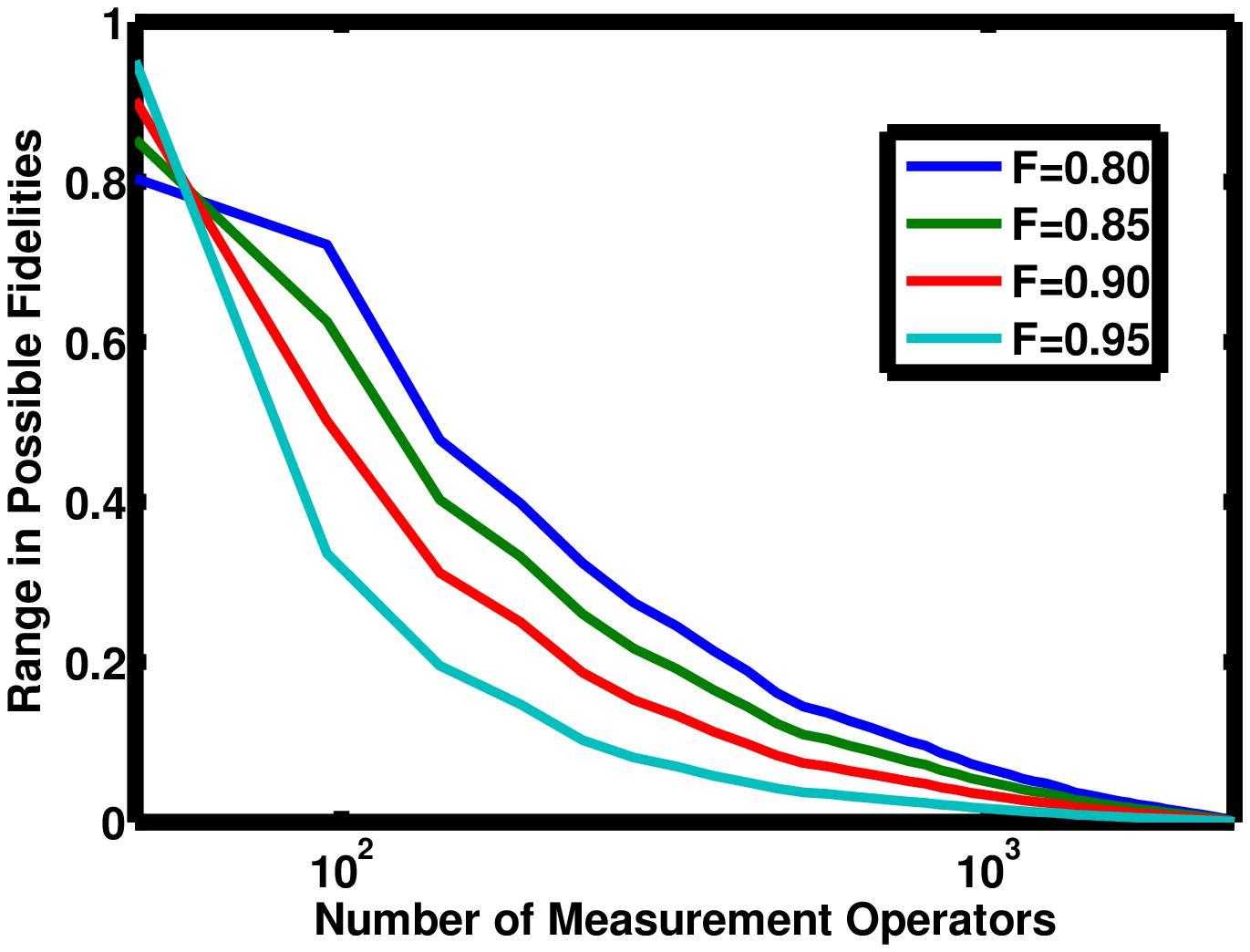}&
\includegraphics[width=8.3cm,clip]{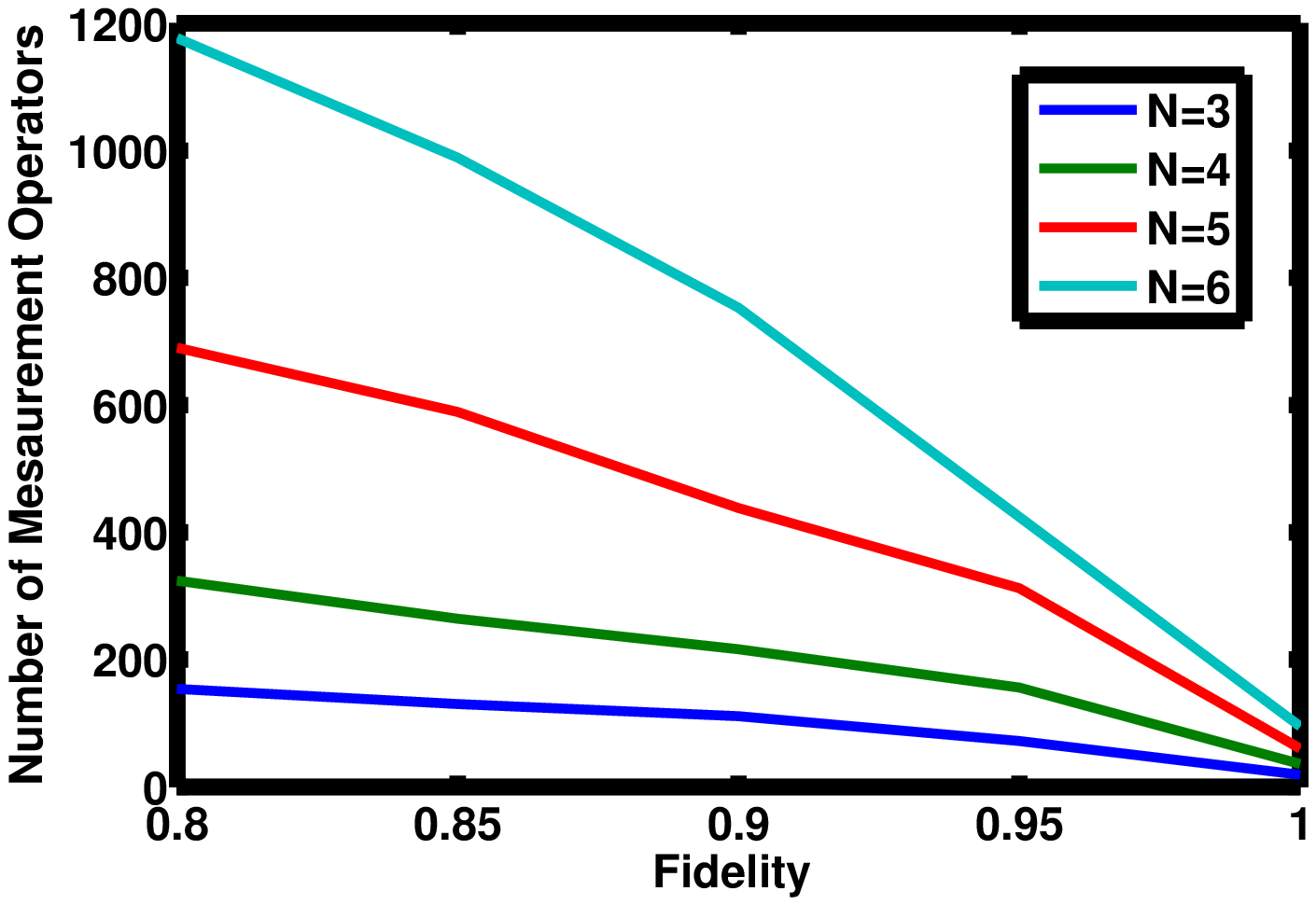}
\end{tabular}
\caption{Numerical analysis of bounding the fidelity of NOON state construction with an incomplete set of measurement outcomes.  All simulations are for noiseless measurements and assume the qubits are initially in their ground states.  (a) Upper and lower bounds for the overlap between the 4-photon NOON state and various mixtures of the this NOON state and the maximally mixed state, plotted vs. the number of randomly sampled measurements performed.  (b)  Lower bounds on the NOON state fidelity for mixtures of different dimension.  The x-axis of this plot is the number of measurement results divided by the dimension of $\mathfrak{su}(d)$ for the different NOON states.  (c)  The difference between the upper and lower bounds for mixtures of the 6-photon NOON state and the maximally mixed state plotted versus the number of measured observables.  (d)  The number of measurements necessary to give a separation of 5\% between the upper and lower bounds, plotted versus the actual fidelity of the experimental state. 
}
\label{F:bounds}
\end{center}
\end{figure*}

To determine the utility of this bounding scheme we ran some simple numerical simulations, the results of which are displayed in Fig.~\ref{F:bounds}.  We place upper and lower bounds on states that are convex sums of the NOON state and the maximally mixed state, or $\rho_{\rm exp} =  p \rho_{\rm NOON} + (1-p) \mathbb{I}/d$.  For a given overlap with the NOON state, these mixtures are generally the states that are hardest to bound, and so represent a worst case test of the protocol.  The measurements were chosen by sampling the $\tau_i$ uniformly from the interval $[0,\pi]$ and the $\alpha$'s uniformly from the interior of the circle on phase space with radius $N$.  For a simple test we looked at NOON states with photon numbers going up to $N=6$, where even the longest semidefinite optimizations were solved in about a minute on a laptop running matlab.

In the case of randomly sampled measurements, the lower bound has a rapid initial rise with respect to the fraction of the possible independent operators measured.  The time for convergence of the upper and lower bounds is highly dependent on the fidelity of the experimental state, with high fidelity states converging very quickly.  In order to get an fairly exact estimate of the fidelity for a density matrix far from the desired state, this method may not be much more efficient than full tomography, however, it is often useful to quickly generate bounds out of small data sets in order to diagnose state generation protocols.  Also, we suspect that this protocol is more amenable to optimizing the measurement operators than full tomography, and so it should be possible to get faster convergence than can be achieved by random sampling.

\section{Conclusions}

In conclusion, we have shown how to determinisitally create large NOON states of microwave photons in superconducting resonators, and how to verify the success of this experiment in a more efficient way than full state tomography. It can be shown, that this reconstruction protocol can be extended to account for detector noise. 

NOON states in the optical domain have been proposed for super-resolution in quantum lithography and interferometry \cite{Boto00,Kok02,Jones09}. With wavelengths of microwave photons in our case being on the order of centimetres, lithograqhy is unlikely for microwave photons. Microwave interferometry has applications in plasma diagnostics \cite{Laroussi99} and astrophysics relating to the cosmic microwave background. The application of NOON states to these systems remains speculative, and will be explored in future work. 

We acknowledge useful discussions with J.M. Martinis and A. Cleland, who independently developed a similar protocol for NOON state creation, as well as discussions with M. Mariantoni, H. Wang, F. Motzoi, and B. McNamara. Work supported by NSERC through QuantumWorks and the discovery grants program and, in parts, by DARPA through the QuEST program and IARPA through the MQCO program.

\bibliography{NOON,frankslibrary,frankspapers}

\begin{thebibliography}{35}
\expandafter\ifx\csname natexlab\endcsname\relax\def\natexlab#1{#1}\fi
\expandafter\ifx\csname bibnamefont\endcsname\relax
  \def\bibnamefont#1{#1}\fi
\expandafter\ifx\csname bibfnamefont\endcsname\relax
  \def\bibfnamefont#1{#1}\fi
\expandafter\ifx\csname citenamefont\endcsname\relax
  \def\citenamefont#1{#1}\fi
\expandafter\ifx\csname url\endcsname\relax
  \def\url#1{\texttt{#1}}\fi
\expandafter\ifx\csname urlprefix\endcsname\relax\def\urlprefix{URL }\fi
\providecommand{\bibinfo}[2]{#2}
\providecommand{\eprint}[2][]{\url{#2}}

\bibitem[{\citenamefont{Glauber}(1963)}]{Glauber63}
\bibinfo{author}{\bibfnamefont{R.}~\bibnamefont{Glauber}},
  \bibinfo{journal}{Phys. Rev.} \textbf{\bibinfo{volume}{131}},
  \bibinfo{pages}{2766} (\bibinfo{year}{1963}).

\bibitem[{\citenamefont{Cahill and Glauber}(1969)}]{Cahill69}
\bibinfo{author}{\bibfnamefont{K.}~\bibnamefont{Cahill}} \bibnamefont{and}
  \bibinfo{author}{\bibfnamefont{R.}~\bibnamefont{Glauber}},
  \bibinfo{journal}{Phys. Rev.} \textbf{\bibinfo{volume}{177}},
  \bibinfo{pages}{1882} (\bibinfo{year}{1969}).

\bibitem[{\citenamefont{Blais et~al.}(2004)\citenamefont{Blais, Huang,
  Wallraff, Girvin, and Schoelkopf}}]{Blais04}
\bibinfo{author}{\bibfnamefont{A.}~\bibnamefont{Blais}},
  \bibinfo{author}{\bibfnamefont{R.-S.} \bibnamefont{Huang}},
  \bibinfo{author}{\bibfnamefont{A.}~\bibnamefont{Wallraff}},
  \bibinfo{author}{\bibfnamefont{S.}~\bibnamefont{Girvin}}, \bibnamefont{and}
  \bibinfo{author}{\bibfnamefont{R.}~\bibnamefont{Schoelkopf}},
  \bibinfo{journal}{Phys. Rev. A} \textbf{\bibinfo{volume}{69}},
  \bibinfo{pages}{062320} (\bibinfo{year}{2004}).

\bibitem[{\citenamefont{Schoelkopf and Girvin}(2008)}]{Schoelkopf08}
\bibinfo{author}{\bibfnamefont{R.}~\bibnamefont{Schoelkopf}} \bibnamefont{and}
  \bibinfo{author}{\bibfnamefont{S.}~\bibnamefont{Girvin}},
  \bibinfo{journal}{Nature} \textbf{\bibinfo{volume}{451}},
  \bibinfo{pages}{664} (\bibinfo{year}{2008}).

\bibitem[{\citenamefont{Clarke and Wilhelm}(2008)}]{Insight}
\bibinfo{author}{\bibfnamefont{J.}~\bibnamefont{Clarke}} \bibnamefont{and}
  \bibinfo{author}{\bibfnamefont{F.}~\bibnamefont{Wilhelm}},
  \bibinfo{journal}{Nature} \textbf{\bibinfo{volume}{453}},
  \bibinfo{pages}{1031} (\bibinfo{year}{2008}).

\bibitem[{\citenamefont{Hofheinz et~al.}(2009)\citenamefont{Hofheinz, Wang,
  Ansmann, Bialczak, Lucero, Neeley, O'Connell, Sank, Wenner, Martinis
  et~al.}}]{Hofheinz09}
\bibinfo{author}{\bibfnamefont{M.}~\bibnamefont{Hofheinz}},
  \bibinfo{author}{\bibfnamefont{H.}~\bibnamefont{Wang}},
  \bibinfo{author}{\bibfnamefont{M.}~\bibnamefont{Ansmann}},
  \bibinfo{author}{\bibfnamefont{R.~C.} \bibnamefont{Bialczak}},
  \bibinfo{author}{\bibfnamefont{E.}~\bibnamefont{Lucero}},
  \bibinfo{author}{\bibfnamefont{M.}~\bibnamefont{Neeley}},
  \bibinfo{author}{\bibfnamefont{A.~D.} \bibnamefont{O'Connell}},
  \bibinfo{author}{\bibfnamefont{D.}~\bibnamefont{Sank}},
  \bibinfo{author}{\bibfnamefont{J.}~\bibnamefont{Wenner}},
  \bibinfo{author}{\bibfnamefont{J.~M.} \bibnamefont{Martinis}},
  \bibnamefont{et~al.}, \bibinfo{journal}{Nature}
  \textbf{\bibinfo{volume}{459}}, \bibinfo{pages}{546} (\bibinfo{year}{2009}).

\bibitem[{\citenamefont{Ansmann et~al.}(2009)\citenamefont{Ansmann, Wang,
  Bialczak, Hofheinz, Lucero, Neeley, O'Connell, Sank, Weides, Wenner
  et~al.}}]{Ansmann09}
\bibinfo{author}{\bibfnamefont{M.}~\bibnamefont{Ansmann}},
  \bibinfo{author}{\bibfnamefont{H.}~\bibnamefont{Wang}},
  \bibinfo{author}{\bibfnamefont{R.~C.} \bibnamefont{Bialczak}},
  \bibinfo{author}{\bibfnamefont{M.}~\bibnamefont{Hofheinz}},
  \bibinfo{author}{\bibfnamefont{E.}~\bibnamefont{Lucero}},
  \bibinfo{author}{\bibfnamefont{M.}~\bibnamefont{Neeley}},
  \bibinfo{author}{\bibfnamefont{A.~D.} \bibnamefont{O'Connell}},
  \bibinfo{author}{\bibfnamefont{D.}~\bibnamefont{Sank}},
  \bibinfo{author}{\bibfnamefont{M.}~\bibnamefont{Weides}},
  \bibinfo{author}{\bibfnamefont{J.}~\bibnamefont{Wenner}},
  \bibnamefont{et~al.}, \bibinfo{journal}{Nature}
  \textbf{\bibinfo{volume}{461}}, \bibinfo{pages}{504} (\bibinfo{year}{2009}).

\bibitem[{\citenamefont{Chow et~al.}()\citenamefont{Chow, DiCarlo, Gambetta,
  Nunnenkamp, Bishop, Frunzio, Devoret, Girvin, and Schoelkopf}}]{Chow09}
\bibinfo{author}{\bibfnamefont{J.}~\bibnamefont{Chow}},
  \bibinfo{author}{\bibfnamefont{L.}~\bibnamefont{DiCarlo}},
  \bibinfo{author}{\bibfnamefont{J.}~\bibnamefont{Gambetta}},
  \bibinfo{author}{\bibfnamefont{A.}~\bibnamefont{Nunnenkamp}},
  \bibinfo{author}{\bibfnamefont{L.}~\bibnamefont{Bishop}},
  \bibinfo{author}{\bibfnamefont{L.}~\bibnamefont{Frunzio}},
  \bibinfo{author}{\bibfnamefont{M.}~\bibnamefont{Devoret}},
  \bibinfo{author}{\bibfnamefont{S.}~\bibnamefont{Girvin}}, \bibnamefont{and}
  \bibinfo{author}{\bibfnamefont{R.}~\bibnamefont{Schoelkopf}},
  \bibinfo{note}{arXiv:0908.1955}.

\bibitem[{\citenamefont{DiCarlo et~al.}(2009)\citenamefont{DiCarlo, Chow,
  Gambetta, Bishop, Johnson, Schuster, Majer, Blais, Frunzio, Girvin
  et~al.}}]{DiCarlo09b}
\bibinfo{author}{\bibfnamefont{L.}~\bibnamefont{DiCarlo}},
  \bibinfo{author}{\bibfnamefont{J.}~\bibnamefont{Chow}},
  \bibinfo{author}{\bibfnamefont{J.}~\bibnamefont{Gambetta}},
  \bibinfo{author}{\bibfnamefont{L.}~\bibnamefont{Bishop}},
  \bibinfo{author}{\bibfnamefont{B.}~\bibnamefont{Johnson}},
  \bibinfo{author}{\bibfnamefont{D.}~\bibnamefont{Schuster}},
  \bibinfo{author}{\bibfnamefont{J.}~\bibnamefont{Majer}},
  \bibinfo{author}{\bibfnamefont{A.}~\bibnamefont{Blais}},
  \bibinfo{author}{\bibfnamefont{L.}~\bibnamefont{Frunzio}},
  \bibinfo{author}{\bibfnamefont{S.}~\bibnamefont{Girvin}},
  \bibnamefont{et~al.}, \bibinfo{journal}{Nature}
  \textbf{\bibinfo{volume}{460}}, \bibinfo{pages}{240} (\bibinfo{year}{2009}).

\bibitem[{\citenamefont{Boto et~al.}(2000)\citenamefont{Boto, Kok, Abrams,
  Brownstein, Williams, and Dowling}}]{Boto00}
\bibinfo{author}{\bibfnamefont{A.}~\bibnamefont{Boto}},
  \bibinfo{author}{\bibfnamefont{P.}~\bibnamefont{Kok}},
  \bibinfo{author}{\bibfnamefont{D.}~\bibnamefont{Abrams}},
  \bibinfo{author}{\bibfnamefont{S.}~\bibnamefont{Brownstein}},
  \bibinfo{author}{\bibfnamefont{C.}~\bibnamefont{Williams}}, \bibnamefont{and}
  \bibinfo{author}{\bibfnamefont{J.}~\bibnamefont{Dowling}},
  \bibinfo{journal}{Phys. Rev. Lett.} \textbf{\bibinfo{volume}{85}},
  \bibinfo{pages}{2733} (\bibinfo{year}{2000}).

\bibitem[{\citenamefont{Kok et~al.}(2002)\citenamefont{Kok, Lee, and
  Dowling}}]{Kok02}
\bibinfo{author}{\bibfnamefont{P.}~\bibnamefont{Kok}},
  \bibinfo{author}{\bibfnamefont{H.}~\bibnamefont{Lee}}, \bibnamefont{and}
  \bibinfo{author}{\bibfnamefont{J.}~\bibnamefont{Dowling}},
  \bibinfo{journal}{Phys. Rev. A} \textbf{\bibinfo{volume}{65}},
  \bibinfo{pages}{052104} (\bibinfo{year}{2002}).

\bibitem[{\citenamefont{Sanders}(1989)}]{Sanders89}
\bibinfo{author}{\bibfnamefont{B.}~\bibnamefont{Sanders}},
  \bibinfo{journal}{Phys. Rev. A} \textbf{\bibinfo{volume}{40}},
  \bibinfo{pages}{2417} (\bibinfo{year}{1989}).

\bibitem[{\citenamefont{Mitchell et~al.}(2004)\citenamefont{Mitchell, Lundeen,
  and Steinberg}}]{Mitchell04}
\bibinfo{author}{\bibfnamefont{M.}~\bibnamefont{Mitchell}},
  \bibinfo{author}{\bibfnamefont{J.}~\bibnamefont{Lundeen}}, \bibnamefont{and}
  \bibinfo{author}{\bibfnamefont{A.}~\bibnamefont{Steinberg}},
  \bibinfo{journal}{Nature} \textbf{\bibinfo{volume}{429}},
  \bibinfo{pages}{161} (\bibinfo{year}{2004}).

\bibitem[{\citenamefont{Nagata et~al.}(2006)\citenamefont{Nagata, Okamoto,
  O'Brien, Sasaki, and Takeuchi}}]{Nagata07}
\bibinfo{author}{\bibfnamefont{T.}~\bibnamefont{Nagata}},
  \bibinfo{author}{\bibfnamefont{R.}~\bibnamefont{Okamoto}},
  \bibinfo{author}{\bibfnamefont{J.}~\bibnamefont{O'Brien}},
  \bibinfo{author}{\bibfnamefont{K.}~\bibnamefont{Sasaki}}, \bibnamefont{and}
  \bibinfo{author}{\bibfnamefont{S.}~\bibnamefont{Takeuchi}},
  \bibinfo{journal}{Science} \textbf{\bibinfo{volume}{316}},
  \bibinfo{pages}{726} (\bibinfo{year}{2006}).

\bibitem[{\citenamefont{Jones et~al.}(2009)\citenamefont{Jones, Karlen,
  Fitzsimons, Adravan, Benjamin, Briggs, and Morton}}]{Jones09}
\bibinfo{author}{\bibfnamefont{J.}~\bibnamefont{Jones}},
  \bibinfo{author}{\bibfnamefont{S.}~\bibnamefont{Karlen}},
  \bibinfo{author}{\bibfnamefont{J.}~\bibnamefont{Fitzsimons}},
  \bibinfo{author}{\bibfnamefont{A.}~\bibnamefont{Adravan}},
  \bibinfo{author}{\bibfnamefont{S.}~\bibnamefont{Benjamin}},
  \bibinfo{author}{\bibfnamefont{G.}~\bibnamefont{Briggs}}, \bibnamefont{and}
  \bibinfo{author}{\bibfnamefont{J.}~\bibnamefont{Morton}},
  \bibinfo{journal}{Science} \textbf{\bibinfo{volume}{324}},
  \bibinfo{pages}{1166} (\bibinfo{year}{2009}).

\bibitem[{\citenamefont{Berkley et~al.}(2003)\citenamefont{Berkley, Xu, Ramos,
  Gubrud, Strauch, Johnson, Anderson, Dragt, Lobb, and Wellstood}}]{Berkley03}
\bibinfo{author}{\bibfnamefont{A.}~\bibnamefont{Berkley}},
  \bibinfo{author}{\bibfnamefont{H.}~\bibnamefont{Xu}},
  \bibinfo{author}{\bibfnamefont{R.}~\bibnamefont{Ramos}},
  \bibinfo{author}{\bibfnamefont{M.}~\bibnamefont{Gubrud}},
  \bibinfo{author}{\bibfnamefont{F.}~\bibnamefont{Strauch}},
  \bibinfo{author}{\bibfnamefont{P.}~\bibnamefont{Johnson}},
  \bibinfo{author}{\bibfnamefont{J.}~\bibnamefont{Anderson}},
  \bibinfo{author}{\bibfnamefont{A.}~\bibnamefont{Dragt}},
  \bibinfo{author}{\bibfnamefont{C.}~\bibnamefont{Lobb}}, \bibnamefont{and}
  \bibinfo{author}{\bibfnamefont{F.}~\bibnamefont{Wellstood}},
  \bibinfo{journal}{Science} \textbf{\bibinfo{volume}{300}},
  \bibinfo{pages}{1548} (\bibinfo{year}{2003}).

\bibitem[{\citenamefont{McDermott et~al.}(2005)\citenamefont{McDermott,
  Simmonds, Steffen, Cooper, Cicak, Osborn, Oh, Pappas, and
  Martinis}}]{McDermott05}
\bibinfo{author}{\bibfnamefont{R.}~\bibnamefont{McDermott}},
  \bibinfo{author}{\bibfnamefont{R.}~\bibnamefont{Simmonds}},
  \bibinfo{author}{\bibfnamefont{M.}~\bibnamefont{Steffen}},
  \bibinfo{author}{\bibfnamefont{K.}~\bibnamefont{Cooper}},
  \bibinfo{author}{\bibfnamefont{K.}~\bibnamefont{Cicak}},
  \bibinfo{author}{\bibfnamefont{K.}~\bibnamefont{Osborn}},
  \bibinfo{author}{\bibfnamefont{S.}~\bibnamefont{Oh}},
  \bibinfo{author}{\bibfnamefont{D.}~\bibnamefont{Pappas}}, \bibnamefont{and}
  \bibinfo{author}{\bibfnamefont{J.}~\bibnamefont{Martinis}},
  \bibinfo{journal}{Science} \textbf{\bibinfo{volume}{307}},
  \bibinfo{pages}{1299} (\bibinfo{year}{2005}).

\bibitem[{\citenamefont{Steffen et~al.}(2006)\citenamefont{Steffen, Ansmann,
  Bialczak, Katz, Lucero, McDermott, Neeley, Weig, Cleland, and
  Martinis}}]{Steffen06}
\bibinfo{author}{\bibfnamefont{M.}~\bibnamefont{Steffen}},
  \bibinfo{author}{\bibfnamefont{M.}~\bibnamefont{Ansmann}},
  \bibinfo{author}{\bibfnamefont{R.~C.} \bibnamefont{Bialczak}},
  \bibinfo{author}{\bibfnamefont{N.}~\bibnamefont{Katz}},
  \bibinfo{author}{\bibfnamefont{E.}~\bibnamefont{Lucero}},
  \bibinfo{author}{\bibfnamefont{R.}~\bibnamefont{McDermott}},
  \bibinfo{author}{\bibfnamefont{M.}~\bibnamefont{Neeley}},
  \bibinfo{author}{\bibfnamefont{E.~M.} \bibnamefont{Weig}},
  \bibinfo{author}{\bibfnamefont{A.~N.} \bibnamefont{Cleland}},
  \bibnamefont{and} \bibinfo{author}{\bibfnamefont{J.}~\bibnamefont{Martinis}},
  \bibinfo{journal}{Science} \textbf{\bibinfo{volume}{313}},
  \bibinfo{pages}{1423} (\bibinfo{year}{2006}).

\bibitem[{\citenamefont{Sillanp\"a\"a et~al.}(2007)\citenamefont{Sillanp\"a\"a,
  Park, and Simmonds}}]{Sillanpaa07}
\bibinfo{author}{\bibfnamefont{M.}~\bibnamefont{Sillanp\"a\"a}},
  \bibinfo{author}{\bibfnamefont{J.}~\bibnamefont{Park}}, \bibnamefont{and}
  \bibinfo{author}{\bibfnamefont{R.}~\bibnamefont{Simmonds}},
  \bibinfo{journal}{Nature} \textbf{\bibinfo{volume}{449}},
  \bibinfo{pages}{438} (\bibinfo{year}{2007}).

\bibitem[{\citenamefont{Majer et~al.}(2007)\citenamefont{Majer, Chow, Gambetta,
  Koch, Johnson, Schreier, Frunzio, Schuster, Houck, Wallraff
  et~al.}}]{Majer07}
\bibinfo{author}{\bibfnamefont{J.}~\bibnamefont{Majer}},
  \bibinfo{author}{\bibfnamefont{J.}~\bibnamefont{Chow}},
  \bibinfo{author}{\bibfnamefont{J.}~\bibnamefont{Gambetta}},
  \bibinfo{author}{\bibfnamefont{J.}~\bibnamefont{Koch}},
  \bibinfo{author}{\bibfnamefont{B.}~\bibnamefont{Johnson}},
  \bibinfo{author}{\bibfnamefont{J.}~\bibnamefont{Schreier}},
  \bibinfo{author}{\bibfnamefont{L.}~\bibnamefont{Frunzio}},
  \bibinfo{author}{\bibfnamefont{D.}~\bibnamefont{Schuster}},
  \bibinfo{author}{\bibfnamefont{A.}~\bibnamefont{Houck}},
  \bibinfo{author}{\bibfnamefont{A.}~\bibnamefont{Wallraff}},
  \bibnamefont{et~al.}, \bibinfo{journal}{Nature}
  \textbf{\bibinfo{volume}{449}}, \bibinfo{pages}{443} (\bibinfo{year}{2007}).

\bibitem[{\citenamefont{Jaynes and Cummings}(1963)}]{Jaynes63}
\bibinfo{author}{\bibfnamefont{E.}~\bibnamefont{Jaynes}} \bibnamefont{and}
  \bibinfo{author}{\bibfnamefont{F.}~\bibnamefont{Cummings}},
  \bibinfo{journal}{Proc. IEEE} \textbf{\bibinfo{volume}{51}},
  \bibinfo{pages}{89} (\bibinfo{year}{1963}).

\bibitem[{\citenamefont{Hofheinz et~al.}(2008)\citenamefont{Hofheinz, Weig,
  Ansmann, Bialczak, Lucero, Neeley, O'Connell, Wang, and
  Martinis}}]{Hofheinz08}
\bibinfo{author}{\bibfnamefont{M.}~\bibnamefont{Hofheinz}},
  \bibinfo{author}{\bibfnamefont{E.}~\bibnamefont{Weig}},
  \bibinfo{author}{\bibfnamefont{M.}~\bibnamefont{Ansmann}},
  \bibinfo{author}{\bibfnamefont{R.}~\bibnamefont{Bialczak}},
  \bibinfo{author}{\bibfnamefont{E.}~\bibnamefont{Lucero}},
  \bibinfo{author}{\bibfnamefont{M.}~\bibnamefont{Neeley}},
  \bibinfo{author}{\bibfnamefont{A.}~\bibnamefont{O'Connell}},
  \bibinfo{author}{\bibfnamefont{H.}~\bibnamefont{Wang}}, \bibnamefont{and}
  \bibinfo{author}{\bibfnamefont{J.}~\bibnamefont{Martinis}},
  \bibinfo{journal}{Nature} \textbf{\bibinfo{volume}{454}},
  \bibinfo{pages}{310} (\bibinfo{year}{2008}).

\bibitem[{\citenamefont{Motzoi et~al.}(2009)\citenamefont{Motzoi, Gambetta,
  Rebentrost, and Wilhelm}}]{Motzoi09}
\bibinfo{author}{\bibfnamefont{F.}~\bibnamefont{Motzoi}},
  \bibinfo{author}{\bibfnamefont{J.}~\bibnamefont{Gambetta}},
  \bibinfo{author}{\bibfnamefont{P.}~\bibnamefont{Rebentrost}},
  \bibnamefont{and} \bibinfo{author}{\bibfnamefont{F.}~\bibnamefont{Wilhelm}},
  \bibinfo{journal}{Phys. Rev. Lett} \textbf{\bibinfo{volume}{103}},
  \bibinfo{pages}{110501} (\bibinfo{year}{2009}).

\bibitem[{\citenamefont{Law and Eberly}(1996)}]{Law96}
\bibinfo{author}{\bibfnamefont{C.~K.} \bibnamefont{Law}} \bibnamefont{and}
  \bibinfo{author}{\bibfnamefont{J.~H.} \bibnamefont{Eberly}},
  \bibinfo{journal}{Phys. Rev. Lett.} \textbf{\bibinfo{volume}{76}},
  \bibinfo{pages}{1055} (\bibinfo{year}{1996}).

\bibitem[{\citenamefont{Fink et~al.}(2008)\citenamefont{Fink, G\"oppl, Baur,
  Bianchetti, Leek, Blais, and Wallraff}}]{Fink08}
\bibinfo{author}{\bibfnamefont{J.}~\bibnamefont{Fink}},
  \bibinfo{author}{\bibfnamefont{M.}~\bibnamefont{G\"oppl}},
  \bibinfo{author}{\bibfnamefont{M.}~\bibnamefont{Baur}},
  \bibinfo{author}{\bibfnamefont{R.}~\bibnamefont{Bianchetti}},
  \bibinfo{author}{\bibfnamefont{P.}~\bibnamefont{Leek}},
  \bibinfo{author}{\bibfnamefont{A.}~\bibnamefont{Blais}}, \bibnamefont{and}
  \bibinfo{author}{\bibfnamefont{A.}~\bibnamefont{Wallraff}},
  \bibinfo{journal}{Nature} \textbf{\bibinfo{volume}{454}},
  \bibinfo{pages}{315} (\bibinfo{year}{2008}).

\bibitem[{\citenamefont{Khaneja et~al.}(2005)\citenamefont{Khaneja, Reiss,
  Kehlet, Schulte-Herbr\"uggen, and Glaser}}]{Khaneja05}
\bibinfo{author}{\bibfnamefont{N.}~\bibnamefont{Khaneja}},
  \bibinfo{author}{\bibfnamefont{T.}~\bibnamefont{Reiss}},
  \bibinfo{author}{\bibfnamefont{C.}~\bibnamefont{Kehlet}},
  \bibinfo{author}{\bibfnamefont{T.}~\bibnamefont{Schulte-Herbr\"uggen}},
  \bibnamefont{and} \bibinfo{author}{\bibfnamefont{S.}~\bibnamefont{Glaser}},
  \bibinfo{journal}{J. Magn. Reson.} \textbf{\bibinfo{volume}{172}},
  \bibinfo{pages}{296} (\bibinfo{year}{2005}).

\bibitem[{\citenamefont{Sp\"orl et~al.}(2007)\citenamefont{Sp\"orl,
  Schulte-Herbrueggen, Glaser, Bergholm, Storcz, Ferber, and
  Wilhelm}}]{Spoerl07}
\bibinfo{author}{\bibfnamefont{A.}~\bibnamefont{Sp\"orl}},
  \bibinfo{author}{\bibfnamefont{T.}~\bibnamefont{Schulte-Herbrueggen}},
  \bibinfo{author}{\bibfnamefont{S.}~\bibnamefont{Glaser}},
  \bibinfo{author}{\bibfnamefont{V.}~\bibnamefont{Bergholm}},
  \bibinfo{author}{\bibfnamefont{M.}~\bibnamefont{Storcz}},
  \bibinfo{author}{\bibfnamefont{J.}~\bibnamefont{Ferber}}, \bibnamefont{and}
  \bibinfo{author}{\bibfnamefont{F.}~\bibnamefont{Wilhelm}},
  \bibinfo{journal}{Phys. Rev. A} \textbf{\bibinfo{volume}{75}},
  \bibinfo{pages}{012302} (\bibinfo{year}{2007}).

\bibitem[{\citenamefont{Martinis}(2009)}]{Martinis09c}
\bibinfo{author}{\bibfnamefont{J.}~\bibnamefont{Martinis}},
  \bibinfo{journal}{Quantum Information Processing}
  \textbf{\bibinfo{volume}{8}}, \bibinfo{pages}{81} (\bibinfo{year}{2009}).

\bibitem[{\citenamefont{Wang et~al.}(2009)\citenamefont{Wang, Hofheinz,
  Ansmann, Bialczak, Lucero, Neeley, O'Connell, Sank, Weides, Wenner
  et~al.}}]{Wang09}
\bibinfo{author}{\bibfnamefont{H.}~\bibnamefont{Wang}},
  \bibinfo{author}{\bibfnamefont{M.}~\bibnamefont{Hofheinz}},
  \bibinfo{author}{\bibfnamefont{M.}~\bibnamefont{Ansmann}},
  \bibinfo{author}{\bibfnamefont{R.}~\bibnamefont{Bialczak}},
  \bibinfo{author}{\bibfnamefont{E.}~\bibnamefont{Lucero}},
  \bibinfo{author}{\bibfnamefont{M.}~\bibnamefont{Neeley}},
  \bibinfo{author}{\bibfnamefont{A.}~\bibnamefont{O'Connell}},
  \bibinfo{author}{\bibfnamefont{D.}~\bibnamefont{Sank}},
  \bibinfo{author}{\bibfnamefont{M.}~\bibnamefont{Weides}},
  \bibinfo{author}{\bibfnamefont{J.}~\bibnamefont{Wenner}},
  \bibnamefont{et~al.}, \bibinfo{journal}{Phys. Rev. Lett.}
  \textbf{\bibinfo{volume}{103}} (\bibinfo{year}{2009}).

\bibitem[{\citenamefont{Koch et~al.}(2007)\citenamefont{Koch, Yu, Gambetta,
  Houck, Schuster, Majer, Blais, Devoret, Girvin, and Schoelkopf}}]{Koch07c}
\bibinfo{author}{\bibfnamefont{J.}~\bibnamefont{Koch}},
  \bibinfo{author}{\bibfnamefont{T.}~\bibnamefont{Yu}},
  \bibinfo{author}{\bibfnamefont{J.}~\bibnamefont{Gambetta}},
  \bibinfo{author}{\bibfnamefont{A.}~\bibnamefont{Houck}},
  \bibinfo{author}{\bibfnamefont{D.}~\bibnamefont{Schuster}},
  \bibinfo{author}{\bibfnamefont{J.}~\bibnamefont{Majer}},
  \bibinfo{author}{\bibfnamefont{A.}~\bibnamefont{Blais}},
  \bibinfo{author}{\bibfnamefont{M.}~\bibnamefont{Devoret}},
  \bibinfo{author}{\bibfnamefont{S.}~\bibnamefont{Girvin}}, \bibnamefont{and}
  \bibinfo{author}{\bibfnamefont{R.}~\bibnamefont{Schoelkopf}},
  \bibinfo{journal}{Phys. Rev. A} \textbf{\bibinfo{volume}{76}},
  \bibinfo{pages}{042319} (\bibinfo{year}{2007}).

\bibitem[{\citenamefont{Houck et~al.}(2009)\citenamefont{Houck, Koch, Devoret,
  Girvin, and Schoelkopf}}]{Houck09}
\bibinfo{author}{\bibfnamefont{A.}~\bibnamefont{Houck}},
  \bibinfo{author}{\bibfnamefont{J.}~\bibnamefont{Koch}},
  \bibinfo{author}{\bibfnamefont{M.}~\bibnamefont{Devoret}},
  \bibinfo{author}{\bibfnamefont{S.}~\bibnamefont{Girvin}}, \bibnamefont{and}
  \bibinfo{author}{\bibfnamefont{R.}~\bibnamefont{Schoelkopf}},
  \bibinfo{journal}{Quant. Inf. Proc.} \textbf{\bibinfo{volume}{8}},
  \bibinfo{pages}{105} (\bibinfo{year}{2009}).

\bibitem[{\citenamefont{Paris and Rehacek}(2004)}]{Paris2004}
\bibinfo{editor}{\bibfnamefont{M.}~\bibnamefont{Paris}} \bibnamefont{and}
  \bibinfo{editor}{\bibfnamefont{J.}~\bibnamefont{Rehacek}}, eds.,
  \emph{\bibinfo{title}{Quantum State Estimation}}
  (\bibinfo{publisher}{Springer}, \bibinfo{year}{2004}).

\bibitem[{\citenamefont{Vandenberghe and Boyd}(1996)}]{vandenberghe96}
\bibinfo{author}{\bibfnamefont{L.}~\bibnamefont{Vandenberghe}}
  \bibnamefont{and} \bibinfo{author}{\bibfnamefont{S.}~\bibnamefont{Boyd}},
  \bibinfo{journal}{SIAM Review} \textbf{\bibinfo{volume}{38}},
  \bibinfo{pages}{49} (\bibinfo{year}{1996}).

\bibitem[{\citenamefont{Merkel et~al.}(2010)\citenamefont{Merkel, Riofr\'\i{}o,
  Flammia, and Deutsch}}]{Merkel2010}
\bibinfo{author}{\bibfnamefont{S.~T.} \bibnamefont{Merkel}},
  \bibinfo{author}{\bibfnamefont{C.~A.} \bibnamefont{Riofr\'\i{}o}},
  \bibinfo{author}{\bibfnamefont{S.~T.} \bibnamefont{Flammia}},
  \bibnamefont{and} \bibinfo{author}{\bibfnamefont{I.~H.}
  \bibnamefont{Deutsch}}, \bibinfo{journal}{Phys. Rev. A}
  \textbf{\bibinfo{volume}{81}}, \bibinfo{pages}{032126}
  (\bibinfo{year}{2010}).

\bibitem[{\citenamefont{Laroussi}(1999)}]{Laroussi99}
\bibinfo{author}{\bibfnamefont{M.}~\bibnamefont{Laroussi}},
  \bibinfo{journal}{Int. J. Infrared and Millimeter Waves}
  \textbf{\bibinfo{volume}{8}}, \bibinfo{pages}{1501} (\bibinfo{year}{1999}).

\end{thebibliography}
\end{document}